\documentstyle[11pt,aaspp4]{article}

\lefthead{Brandt et al.}
\righthead{X-Ray Background}

\def\simgt{\lower 2pt \hbox{$\, \buildrel {\scriptstyle >}\over {\scriptstyle
\sim}\,$}}
\def\simlt{\lower 2pt \hbox{$\, \buildrel {\scriptstyle <}\over {\scriptstyle
\sim}\,$}}

\def\asca{{\it ASCA\/}}

\def\chandra{{\it Chandra\/}}

\def\heao1{{\it HEAO-1\/}}

\def\rosat{{\it ROSAT\/}}

\def\sax{{\it BeppoSAX\/}}

\def\vela5b{{\it Vela-5B\/}}

\def\crsscluster{CRSS~J0030.5+2618}
\def\crssqso{CRSS~J0030.6+2620}
\def\crsssy2{CRSS~J0030.7+2616}
\begin{document}

\title{Observations of Faint, Hard-Band X-ray Sources 
in the Field of \crsscluster\ with the Chandra X-ray Observatory 
and the Hobby-Eberly Telescope\footnote{Based on 
observations obtained with the Hobby-Eberly
Telescope, which is a joint project of the University of Texas at Austin,
the Pennsylvania State University, Stanford University,
Ludwig-Maximillians-Universit\"at M\"unchen, and Georg-August-Universit\"at
G\"ottingen.}
}


\author{
W.N.~Brandt\altaffilmark{\ref{PennState}},
A.E.~Hornschemeier\altaffilmark{\ref{PennState}},
D.P.~Schneider\altaffilmark{\ref{PennState}},
G.P.~Garmire\altaffilmark{\ref{PennState}},
G.~Chartas\altaffilmark{\ref{PennState}},
Gary~J.~Hill\altaffilmark{\ref{Texas}},
P.J.~MacQueen\altaffilmark{\ref{Texas}}, 
L.K.~Townsley\altaffilmark{\ref{PennState}},
D.N.~Burrows\altaffilmark{\ref{PennState}},
T.S.~Koch\altaffilmark{\ref{PennState}},
J.A.~Nousek\altaffilmark{\ref{PennState}} and
L.W.~Ramsey\altaffilmark{\ref{PennState}}
}


\newcounter{address}
\setcounter{address}{2}
\altaffiltext{\theaddress}{Department of Astronomy and Astrophysics,
The Pennsylvania State University, University Park, PA 16802
\label{PennState}}
\addtocounter{address}{1}
\altaffiltext{\theaddress}{McDonald Observatory,
University of Texas, Austin, TX 78712
\label{Texas}}
\addtocounter{address}{1}


\vbox{
\begin{abstract}
We present results from a study of 2--8~keV X-ray sources detected by
the Advanced CCD Imaging Spectrometer (ACIS) instrument 
on the \chandra\ X-ray Observatory
in the field of the $z=0.516$ cluster \crsscluster. 
In our 63.5~arcmin$^2$ search area, we detect 10 sources with 
2--8~keV fluxes down to $\approx 4\times 10^{-15}$~erg~cm$^{-2}$~s$^{-1}$;
our lowest flux sources are $\approx 10$ times fainter than those 
previously available for study in this band. Our derived source
density is about an order of magnitude larger than previous source 
counts above 2~keV, although this density may be enhanced somewhat 
due to the presence of the cluster. 
We detail our methods for source detection and characterization,
and we show that the resulting source list and parameters are robust. 
We have used the Marcario Low Resolution Spectrograph on the Hobby-Eberly 
Telescope to obtain optical spectra for several of our sources; combining 
these spectra with archival data we find that the sources 
appear to be active galaxies, often with
narrow permitted lines, red optical continua or hard X-ray spectra. Four of 
the X-ray sources are undetected to $R=21.7$; if they reside in 
$L^\star$ galaxies they must have $z>$~0.55--0.75 and hard X-ray 
luminosities of $L_{2-8}\simgt 4\times 10^{42}$~erg~s$^{-1}$. 
We detect all but one of our 2--8~keV sources in 
the 0.2--2~keV band as well. This result extends to 
significantly lower fluxes the constraints on any large, completely new 
population of X-ray sources that appears above 2--3~keV. 
\end{abstract}
}



%


\section{Introduction}

About 70\% of the extragalactic 
X-ray background in the soft 0.5--2~keV band has been 
resolved into discrete sources by pencil-beam surveys with the \rosat\ 
satellite (e.g., \cite{has98}). The 0.5--2~keV 
source counts have reached a surface 
density of $\approx 1000$~deg$^{-2}$ at a discrete source 
detection limit of $\approx 10^{-15}$~erg~cm$^{-2}$~s$^{-1}$,
and simple extrapolations suggest that all of the 0.5--2~keV 
extragalactic X-ray background will be resolved by a flux limit of 
$\sim 2\times 10^{-16}$~erg~cm$^{-2}$~s$^{-1}$
at which the surface density will be $\sim 3000$~deg$^{-2}$.
Optical identification programs (e.g., \cite{schmidt98})
have established that type~1 Active Galactic Nuclei (AGN),
such as Seyfert~1 galaxies and Quasi-Stellar Objects (QSOs), 
are the dominant contributors above a 0.5--2~keV flux of 
$\approx 5\times 10^{-15}$~erg~cm$^{-2}$~s$^{-1}$. 
A non-negligible number (about 16\%) of type~2 AGN are
seen as well.  

Largely due to instrumental limitations, the nature of the sources
that produce the $>2$~keV X-ray background is much less certain
at present. It is important to solve this mystery since 
most of the energy density in the X-ray background is located above 
the \rosat\ band. The best current constraints on the sources of
the 2--10~keV X-ray background have come from the \asca\ and \sax\
satellites. Long observations with these satellites have reached 
discrete source detection limits of 
$\approx$~(3--5)$\times 10^{-14}$~erg~cm$^{-2}$~s$^{-1}$
and have resolved $\approx 30$\% of the 2--10~keV background
into discrete sources (e.g., Ogasaka et~al. 1998;
Giommi, Fiore \& Perri 1999). The integrated number of sources, $N$, 
is consistent with the law $N(>S)\propto S^{-3/2}$ 
expected for a uniform distribution of sources in Euclidean space.
The deepest 2--10~keV source counts to date have
resolved $\approx 60$ sources deg$^{-2}$.
The faintest 2--10~keV sources appear to have 
flatter spectra (with energy indices of $\alpha\approx 0.5\pm 0.2$) 
than those of typical unabsorbed AGN (e.g., Ueda et~al. 1998), 
suggesting that a population of sources with spectra similar to that 
of the integrated X-ray background dominates above 2~keV. The population
is thought to be at least partially composed of obscured AGN, and some
of these hard sources have indeed been associated with such 
objects (e.g., Fiore et~al. 1999; Akiyama et~al. 2000). An
important result, however, is that the majority of the hard sources
found thus far appear to have counterparts in the soft X-ray 
band (Giommi, Fiore \& Perri 1998). Obscured AGN might still
create soft X-ray emission via electron-scattered X-rays or
due to non-nuclear X-ray emission (e.g. starburst activity). 

The arcsecond imaging quality and high-energy sensitivity of
the \chandra\ X-ray Observatory (\cite{wod96}) promises to 
revolutionize our understanding of the X-ray background above 2~keV. 
The source confusion and misidentification problems that have
dogged earlier hard X-ray surveys will be eliminated. 
In this paper, we use data from \chandra, the 8-m class Hobby-Eberly 
Telescope (HET), and public archives to study a small but well-defined 
sample of faint X-ray sources in the 2--8~keV band. Our sample contains 
several of the faintest $>2$~keV sources yet detected and identified. 
We address (1) the number counts at faint hard X-ray fluxes,
(2) the nature of the faint hard X-ray sources, and (3) the issue of 
whether most faint hard X-ray sources have soft X-ray counterparts. 
Our X-ray sources
lie in the PG~0027+260 (an eclipsing cataclysmic variable) field
of the Cambridge-Cambridge \rosat\ Serendipity Survey 
(e.g., Boyle, Wilkes \& Elvis 1997). This field contains 
\crsscluster, a $z=0.516$ cluster of galaxies, and it was
observed by \chandra\ for 44~ks during its first month of 
calibration-phase operations. 
The Galactic column density along this line of sight is
$(3.9\pm 0.4)\times 10^{20}$~cm$^{-2}$ (Stark et~al. 1992),
corresponding to an optical depth of $\tau<0.02$ for the
2--8~keV band of primary interest here. 
In this paper we assume 
$H_0=70$~km~s$^{-1}$ Mpc$^{-1}$ and $q_0=\frac{1}{2}$.


\section{X-ray Observations and Data Analysis}

\subsection{ACIS Observation Details and Image Creation}

The field containing \crsscluster\ was observed with the \chandra\ Advanced 
CCD Imaging Spectrometer (ACIS; \cite{garnou99a}; Garmire et~al. 2000, in preparation) 
for a total exposure time of 44~ks on 1999~August~17. 
ACIS consists of ten CCDs designed for efficient
X-ray detection and spectroscopy. Four of the CCDs 
(ACIS-I; CCDs I0--I3) are arranged in a
$2\times 2$ array with each CCD tipped slightly to approximate the
curved focal surface of the \chandra\ High Resolution Mirror Assembly (HRMA). 
The remaining six CCDs (ACIS-S; CCDs S0--S5) are set in a linear array 
and tipped to approximate the Rowland circle of the objective gratings that 
can be inserted behind the HRMA. The CCD which lies on-axis in ACIS-S (S3) 
is orthogonal to the HRMA optical axis. It is a back-illuminated 
CCD that is sensitive for imaging soft X-ray objects.  
Each CCD subtends \hbox{an $8.3^{\prime}\times 8.3^{\prime}$} square on 
the sky. The individual pixels of the CCDs subtend 
$\approx 0.5^{\prime\prime}\times 0.5^{\prime\prime} $ on the sky.
The on-axis image quality of the telescope is approximately~$0.5^{\prime\prime}$ 
(FWHM); this quantity increases to~$\approx 1.0^{\prime\prime}$ (critical sampling 
on the detector) at an off-axis angle of~$\approx 2^{\prime}$. The image size also 
has a weak energy dependence, with poorer quality at higher energy. 

The observation was performed in two segments (observation ID 
numbers 1190 and 1226), separated by 1.0~ks. 
\crsscluster\ was placed at the aim point for the ACIS-S array 
(on CCD S3) during the observation. The aim point position was 
$\alpha_{2000}=00^{\rm h} 30^{\rm m} 32.5^{\rm s}$,
$\delta_{2000}=+26^\circ 18^\prime 13.4^{\prime\prime}$. 
The focal plane temperature was $-99.3^\circ$C.
Faint mode was used for the event telemetry format, and \asca\ 
grade~7 events were rejected on orbit to prevent telemetry
saturation (see \S5.7 of the {\it AXAF Observatory Guide\/} for
a discussion of grades).
Only one 3.3~s frame was `dropped' from the telemetry.  

Here we will focus on the data from CCD S3 since it has not 
shown the charge transfer inefficiency (CTI) increase that has affected
the front-illuminated CCDs (see \cite{garnou99b}). 
To avoid problems associated with the dither of \chandra\
(see \S4.9.2 of the {\it AXAF Observatory Guide\/}), we 
also neglect data within $20^{\prime\prime}$ of the edge of S3. 
Our search area thus comprises 63.5 square arcminutes or
92\% of S3. The two observation segments were co-added 
using the {\sc event browser} software (\cite{broos99}). We 
used the CIAO {\sc datamodel} software, provided by the \chandra\ X-ray 
Center, to create $\approx 0.5^{\prime\prime}$~pixel$^{-1}$ images in the
`full' (0.2--8~keV), 
`soft' (0.2--2~keV), and
`hard' (2--8~keV) bands
(neglecting the 8--10~keV data improves the signal-to-noise 
ratio in the hard band; e.g., Baganoff 1999). 
Our 0.2, 2, and 8~keV band boundaries have uncertainties of
80~eV, 20~eV and 160~eV, respectively. These uncertainties are
smaller than or comparable to the S3 spectral resolution, and 
the 0.2 and 8~keV uncertainties are innocuous due to the small 
effective area of HRMA/ACIS below 0.3~keV and above 8~keV.
The 2~keV band boundary is furthermore convenient because it is
close to the energy of the HRMA response drop due to the
iridium M-edge. 

We have only used events with ACIS grades of 0, 2, 8, 16 and 64. 
For the background level during this observation, this 
conservative grade set appears to 
provide the best overall performance when trying 
to detect faint, hard sources on S3, although we explore other grade 
set choices in \S2.4. For this 44~ks observation, the average  
background in the hard band varies across S3 in the 
range $\approx$~0.03--0.05~count~pixel$^{-1}$. 

We corrected the \chandra\ astrometry by comparison with 
\rosat\ HRI, Palomar Optical Sky Survey (POSS), and 
Isaac Newton Telescope data (see below). The QSO 
\crssqso\ ($z=0.493$) and the Seyfert~2 \crsssy2\ ($z=0.247$) were 
particularly useful in this regard (see Boyle et~al. 1995 and
Boyle et~al. 1997). 

\subsection{Source Searching}

We used the CIAO {\sc wavdetect} software (Dobrzycki et~al. 1999;
Freeman et~al. 2000) to search our 
images for sources. Our primary interest is in 2--8~keV 
point sources. We used a significance threshold of $1\times 10^{-6}$ 
and computed 5 scaled transforms for wavelet scale sizes 
of 1, 2, 4, 6 and 8 pixels. In our hard band image, we have detected 
9 point sources on S3.\footnote{We also detect a source at
$\alpha_{2000}=00^{\rm h} 30^{\rm m} 57.8^{\rm s}$,
$\delta_{2000}=+26^\circ 17^\prime 44.3^{\prime\prime}$, but this
source lies $\approx 19^{\prime\prime}$ from the edge of S3
and is hence excluded from consideration.}
These are listed in Table~1 and shown in Figures~\ref{fig1} 
and \ref{fig2}. 
We estimate that our absolute positions in the hard band are good to 
within $\approx 3^{\prime\prime}$. They are therefore of 
comparable quality to those used in the \rosat\ High-Resolution Imager 
survey of the Lockman Hole (Hasinger et~al. 1998), and they are
{\it much\/} better than earlier positions in the hard X-ray band. 
All of these sources, except source~5, were also detected in 
the independent soft-band image (usually with a significantly higher
number of counts), giving us confidence in their reality.
The relative positional agreement between hard-band and soft-band sources
was typically better than $1^{\prime\prime}$. Given this relative
positional agreement, the probability
that any given source is a false match is $\approx 0.2$\%.  
While the background in the soft band has significant spatial
structure due to instrumental effects (e.g., node boundaries) and
the presence of the cluster, this does not appear to affect our
matching of hard and soft sources. Source~7 lies at a position where
the soft-band background is elevated by $\approx 25$\% by the 
central node boundary. This elevated background arises due to
cosmic rays interacting with the physically separated frame 
store regions of the S3 CCD, and it is blurred when the 
\chandra\ dither is removed by the pipeline software. 
The slightly elevated background does not 
compromise the detection of source~7, but our soft-band 
photometry may have a small systematic error in addition to the
statistical error given in Table~1. In addition, we 
are confident that none of our sources is due to a `hot pixel' because 
these would show the characteristic Lissajous dither pattern from 
the spacecraft aspect.
We have compared the angular extents of our sources with the 90\%
encircled energy radius of the \chandra\ PSF
(see \S3.8 of the {\it AXAF Observatory Guide\/}), and 
although the photon statistics are limited
we find no clear evidence for anomalous extent. 

We have compared our number counts in the soft band with those of
Hasinger et~al. (1998) as a rough consistency check. Hasinger et~al. (1998) 
found $940\pm 170$~sources~deg$^{-2}$ at a 0.5--2~keV flux level
of $1\times 10^{-15}$~erg~cm$^{-2}$~s$^{-1}$. Our source density in
the soft band is $1500\pm 300$~sources~deg$^{-2}$ at a 0.2--2~keV 
flux level of $\approx 6\times 10^{-16}$~erg~cm$^{-2}$~s$^{-1}$.
For plausible bandpass corrections, our number counts are roughly
consistent with a simple extrapolation of the Hasinger et~al. (1998) 
$\log N$--$\log S$ relation. We note that there is probably an
enhancement in our number counts due to the presence of the
cluster \crsscluster\ (see \S4 for further discussion). 

The minimum detectable 2--8~keV flux varies across S3 due to
effects such as point spread function (PSF) broadening, 
vignetting, and spatially dependent CTI. Within 
$3.5^\prime$ of the aim point, 
we estimate our flux limit to be 
(4--5)$\times 10^{-15}$~erg~cm$^{-2}$~s$^{-1}$ (corresponding to
$\approx 6$ counts), while at larger off-axis angles our flux
limit increases fairly quickly (e.g., see Figure~6.5 of the
{\it AXAF Proposers' Guide\/}). While this spatially dependent flux
limit should be kept in mind, none of our main results below 
is sensitive to the precise details of our flux limit. Even at 
the locations on S3 furthest from the aim point, the observation is
$\simgt 5$ times deeper than previous hard X-ray surveys. 

\subsection{Source Parameterization}

{\sc wavdetect} performs photometry on detected sources, and 
we have cross checked the {\sc wavdetect} results with
manual aperture photometry. We find good agreement between 
the two techniques for all 
sources other than source~2, where the {\sc wavdetect} photometry
clearly has failed ({\sc wavdetect} finds source~2 but claims 
it only has 0.9 counts in the
hard band). In Table~1 we quote our manual photometry results for 
source~2 and {\sc wavdetect} photometry results for all 
other sources. These have not been corrected for vignetting. 
We also quote the `band ratio' defined as the ratio of hard-band to 
soft-band counts. The errors for our band ratio values have
been computed following the `numerical method' described in 
\S1.7.3 of Lyons (1991); in the Poisson limit this method is more 
reliable than the standard approximate variance formula
(e.g., see \S3.3 of Bevington \& Robinson 1992). 
In Figure~3 we compare our band ratios to power-law models
with varying amounts of neutral absorption. Several of our
sources appear likely to suffer significant internal
X-ray absorption. 
In addition, {\sc wavdetect} reports a significance level for
each source, defined as the number of source counts divided by
the Gehrels (1986) standard deviation of the number of
background counts (see Table~1). For source~2 we
have calculated the 2--8~keV significance using 
our manual photometry. We do not report a 0.2--2~keV 
significance for source~5 because it is not detected in this 
band. 

We have used {\sc event browser} to create full-band light curves 
for our sources. We analyzed these for variability using 
a Kolmogorov-Smirnov test. Most of our sources do not show 
significant evidence for variability. Source~8 may show
variability by a factor of $\approx 2$ on a timescale of
$\approx 10000$~s. The fact that the photon arrival times
for our sources are distributed fairly evenly throughout the 
observation length is a further argument against some brief, 
transient effect (e.g., a transient `hot pixel' or a cosmic ray) 
producing spurious sources. We have also examined the energy and 
ACIS grade distributions for our sources and find no anomalies 
that might indicate spurious instrumental effects. 

We have calculated vignetting-corrected
0.2--2~keV and 2--8~keV fluxes for our sources 
using the counts from Table~1, and these are given in Table~2.
We assume a $\Gamma=1.9$ power-law model with the Galactic column 
density, and our fluxes have not been corrected for Galactic
or internal absorption. Our 2--8~keV observed fluxes (those of
primary scientific interest here) are quite insensitive to the
assumed column density for $N_{\rm H}\simlt 10^{22}$~cm$^{-2}$
(compare with Figure~3). Our 0.2--2~keV fluxes are somewhat
more sensitive to the assumed column density; for our sources
with large band ratios the 0.2--2~keV fluxes calculated with 
Galactic absorption are 5--20\% lower than those calculated using
$\Gamma=1.9$ and column densities estimated from Figure~3. 
For our flux calculations, we have used the calibration-phase ACIS 
redistribution matrix files (rmfs) from the ACIS calibration group
(S.~Buczkowski \& N.~Schulz, private communication), and
we have also used the calibration-phase ancillary response
files (arfs; N. Schulz, private communication). These spectral
responses are for a focal plane temperature of $-100^\circ$C, and
they assume filtering upon \asca\ grades 0, 2, 3, 4~and 6. To correct
for our more conservative choice of grades, we have multiplied
our 0.2--2~keV fluxes by a factor of $1.23\pm 0.12$ and our 2--8~keV 
fluxes by a factor of $1.53\pm 0.23$ (statistical errors only). 
These factors have been determined by comparing 
the numbers of events for our sources obtained
with the two different grade filtering methods
(source~6 is excluded in these comparisons since it would
otherwise dominate the results; we compute separate factors
for source~6 of $1.39\pm 0.06$ and $1.80\pm 0.16$). We estimate
that our fluxes have calibration uncertainties of $\sim 30$\%, but it 
is clear that we are detecting 2--8~keV sources {\it much\/} fainter
than were detected with \asca\ and \sax. 

In addition to the sources described above, we will introduce
two new sources below: 
source~AG1 (found on S3 when \asca\ grade filtering is used; see \S2.4)
and
source~I3 (found on ACIS CCD I3; see \S3.2). 
To compute fluxes for source~AG1, we have followed
the method of the previous paragraph but have not made the grade 
correction (since we use \asca\ grade filtering for this
source). 
To compute fluxes for source~I3, we again
followed the method of the previous paragraph. However, ACIS 
I3 spectral responses are only available at present
for a focal plane temperature of $-90^\circ$C. We have used
these but recognize that this may introduce systematic error into
our flux calculations for source~I3. Therefore, we do not use 
the fluxes for source~I3 in any of our subsequent analysis. 

\subsection{Additional Safety Checks}

The background level in the S3 CCD shows significant flaring during
the observation due to `space weather' (primarily soft electrons
interacting with the CCD). We have repeated the analysis above 
after editing out the 8.0~ks when the background level was 
highest. We find the same 2--8~keV sources as those listed in Table~1 
except that source~4 is not detected in the edited data set. Source~4 is
detected in the 0.2--2~keV edited data, and we therefore believe that it is 
reliably detected in the hard band in the unedited data set (see Figure~2). 

We have also performed source searches on images where we relax our
grade screening so that we accept \asca\ grades 0, 2, 3, 4 and 6. 
In these searches we detect 
most of the sources discussed in \S2.2, but we fail to 
detect sources 2, 4 and 5 in the hard band. We thus infer a generally
lower source detection efficiency with this grade screening. Our
average background across S3 with this grade screening varies from 
$\approx$~0.07--0.09~count~pixel$^{-1}$, a factor of $\approx 2$ higher 
than in \S2.1 and \S2.2. However, we 
do detect one new hard band source, which we will hereafter 
refer to as `source~AG1' (`AG' is for `\asca\ grade'). We consider this 
source to be reliable since it is also detected in our independent
soft band image, and we give its properties in Tables~1 and 2
(also see Figure~1). This
source appears to have been missed by our source searching in 
\S2.2 because several of its counts were rejected by our conservative
grade filtering prescription. This highlights that it is difficult
to choose a single optimal grade filtering criterion when dealing with
sources with few counts; chance fluctuations in source
grades can be important in this limit. 

We have investigated if our choice of {\sc wavdetect} wavelet scale
sizes affects our results, and we find no evidence for this. We have
repeated the searching of \S2.2 using wavelet scale sizes of
1, 1.414, 2, 2.828, 4, 5.657, 8, 11.314 and 16 pixels 
(a `$\sqrt 2$ sequence'), and we find the same sources as in \S2.2. 

We are developing a matched filter code, based on the HRMA PSF, 
which we have used to check our {\sc wavdetect} source detections. 
This preliminary matched filter code finds the same 9 hard-band sources on S3 
that we have discussed in \S2.1 and \S2.2, and, 
like {\sc wavdetect}, it finds soft-band 
counterparts for all sources other than source~5 (see \S2.2).  

We have also examined the spatial distribution of 2--8~keV sources on
S3 to see if we can detect any spatial non-uniformity. We have used
a two-dimensional `Kolmogorov-Smirnov test' (see \S14.7 of Press et~al. 1992),
and we have performed Monte-Carlo simulations to compute significance
values for small numbers of sources. The 9 hard-band sources of \S2.2 
are found to be consistent with a uniform distribution. We have also
performed the test including source~AG1 (see above), and we found
this sample of 10 sources to be consistent with a uniform distribution. 
However, we note that the two-dimensional `Kolmogorov-Smirnov test'
has limited statistical power for only 9--10 sources. In fact, we might
have expected some spatial nonuniformity due to the fact that our
sensitivity decreases away from the aim point; this may partially 
explain the absence of sources toward the lower-left part of Figure~1. 

We have searched for spatial correspondences between our hard-band source
positions and instrumental features, and we find none. In particular, we
stress that sources 2, 3, 5 and 7 (the four blank-field sources of \S3)
are {\it not\/} linearly distributed 
along the dithered central node boundary (see \S2.2 for a discussion
of the node boundary). Source~7 is the closest of these four to 
the node boundary, and it is still $9^{\prime\prime}$ from it (much 
larger than the PSF size at this position). 


\section{Optical Observations and Data Analysis}

\subsection{Source Matching and Optical Photometry}

We have compared the positions of our 2--8~keV sources with optical 
sources on the Palomar Optical Sky Survey (POSS) 
plates and an archival 600~s $R$-band image taken 
with the 2.5-m Isaac Newton Telescope (INT) on 19 October 1995 (see Figure~4 
and \S2.5 of Boyle et~al. 1997). The INT image is sensitive down to
$R\approx 21.7$. We take an optical source to be positionally 
coincident with a \chandra\ source when its centroid is within
$3^{\prime\prime}$ of the \chandra\ position in Table~1. 
Five of our nine sources from \S2.2 are detected 
on the POSS plates. Two of these are AGN that have been previously 
identified by Boyle et~al. (1997): 
the QSO \crssqso\ at $z=0.493$ (our source~6) and 
the Seyfert~2 \crsssy2\ at $z=0.247$ (our source~8). 
The other four sources are not detected either 
on the POSS plates or in the deeper INT image, and these 
are henceforth referred to as `blank-field sources.'
The number of blank-field sources we have obtained appears 
reasonable when compared with an extrapolation to lower hard 
X-ray fluxes of the $R$-magnitude versus 2--10~keV flux 
relation shown in Figure~3 of Akiyama et~al. (2000); we would
expect optical counterparts with $R\approx$~18--23.5. It is
also generally consistent with the results of deep soft X-ray 
surveys (compare with Figure~3 of Hasinger et~al. 1999). 
We have used the APM catalog (\cite{mi92}) and the INT image
to determine $R$ magnitudes or $R$-magnitude limits for our
sources; these are given in Table~2. 

Source~AG1 from \S2.4 is not detected on the POSS plates, but it
is coincident with a faint ($R=21.5$) object seen in the INT image. 

Using the INT image, we find an $R$-band 
source density of 10.3 per square 
arcminute down to $R=21.7$. Given this source density and our 
$3^{\prime\prime}$ error circles, the probability that any given 
2--8~keV source has a false optical counterpart is 0.08. 
However, we note that most of the counterparts have $R$ magnitudes
that are substantially brighter than the detection limit for the
INT image, and their identifications are correspondingly more
secure. The HET spectroscopy below combined with the rarity of AGN 
on the sky supports the correctness of our optical matching 
(see \S4.1 of Schmidt et~al. 1998 for details). 

We have searched for NRAO VLA Sky Survey (NVSS; Condon et~al. 1998) 
sources coincident with our X-ray sources and find none. This area
of sky has not been covered by the VLA FIRST survey 
(Becker, White \& Helfand 1995). 

\subsection{Hobby-Eberly Telescope Spectroscopy}

We used the HET to obtain spectra for the three unidentified 2--8~keV 
sources from \S2.2 with optical counterparts on the POSS plates. We also 
obtained a spectrum of the $R=19.1$ optical counterpart to a 2--8~keV source 
located on ACIS-I CCD I3 (see Table~1 for source details). We have 
not yet attempted to obtain an optical spectrum for source~AG1. 

The HET, located at McDonald Observatory, is the first optical/infrared 8-m 
class telescope to employ a fixed-altitude (Arecibo-type) design (\cite{lwr98}). 
All spectra were obtained in October~1999 
with the Marcario Low Resolution Spectrograph
(LRS; Hill et~al. 1998; Hill et~al. 2000; 
Schneider et~al. 2000) mounted at the prime focus of the HET. 
A $2.0^{\prime\prime}$ slit and 300~line~mm$^{-1}$ grism/GG385 blocking filter 
produced spectra from 4400~\AA\ to 9000~\AA\ at 24~\AA\ resolution. The exposure
time per source ranged from 20--30 minutes. The image quality as delivered
on the detector was typically $2.5^{\prime\prime}$ (FWHM). Wavelength calibration 
was performed with a fourth-order polynomial fit to a set of Cd/Hg/Ne/Zn lines;
the rms of the fit was 0.8~\AA. Observations of the spectrophotometric
standards of Oke \& Gunn (1983) were used to perform the relative flux 
calibration. Spectra of the four objects are displayed in Figure~5. 

{\bf Source 1:}
Source~1 has H$\alpha$ and [O~{\sc iii}] emission at $z=0.269$, 
with a derived absolute magnitude of $M_{\rm B}=-21.9$. 
It has a large Balmer decrement with 
$H\alpha/H\beta\simgt 10$, and its optical continuum slope is red 
(for an AGN) with 
$\alpha=-2.5\pm 0.4$.\footnote{$F(\nu)\propto \nu^\alpha$. 
Typical `blue' quasars have $-1.3\simlt\alpha\simlt+0.1$.
Optical continuum slopes in this paper are for 5500--8800~\AA\ in the 
observed frame.} The H$\alpha$ line is resolved 
with a FWHM of 1400~km~s$^{-1}$. 

{\bf Source 4:}
Source~4 is definitely a $z=0.247$ galaxy (H$\alpha$, [O~{\sc iii}], and 
[O~{\sc ii}] emission, plus a strong Mg~b absorption feature) with
$M_{\rm B}=-21.3$. Unfortunately, our spectrum did not permit a search
for [Ne~{\sc v}] emission at 3426~\AA\ (compare with \S4 of
Schmidt et~al. 1998). Its Balmer decrement is $H\alpha/H\beta\simgt 3$,
and its optical continuum slope is $\alpha=-1.8\pm 0.4$. The optical
continuum emission is dominated by star light. The H$\alpha$ line
is unresolved with a FWHM of $<900$~km~s$^{-1}$. 
With a 2--8~keV flux of $4.1\times 10^{-15}$~erg~cm$^{-2}$~s$^{-1}$, 
this is the faintest 2--8~keV source yet identified to our knowledge. 
Somewhat surprisingly, the redshift of this source is the same as that 
of \crsssy2\ (our source~8), but we do not have any reason to suspect 
identification problems. A comparison of our HET spectrum 
(see Figure~5) and the spectrum for \crsssy2\ given in Figure~1
of Boyle et~al. (1995) shows that the equivalent width of 
H$\alpha$ in \crsssy2\ is $\simgt 2$ times larger than that of 
source~4. 

{\bf Source 9:}
Source~9 is difficult to interpret. The brightest optical source in the
X-ray error circle is clearly extended on the INT image (see Figure~4).
The optical counterpart is off the X-ray position by 
$\approx 3^{\prime\prime}$, which is by far the largest discrepancy 
of any of the optical identifications shown in Figure~4. 
The HET spectrum for this source shows one strong 
narrow line at 7411~\AA\ that is most likely H$\alpha$ at $z=0.129$. 
The line is unresolved with a FWHM of $<900$~km~s$^{-1}$, and the
optical continuum is red with $\alpha=-2.9\pm 0.4$. The
residual ripple in the continuum below 7000~\AA\ is an artifact due to
incomplete cancellation of fringing from a pellicle by the flat field.
The relatively large difference between the optical and X-ray positions
suggests that this may not be the correct identification. Another possibility
is that this galaxy is a member of a small group and that the X-ray emission
arises from the general environment and not an individual galaxy. However,
the hard X-ray spectral shape would be difficult to understand in this
case. 

{\bf Source~I3:}
Source~I3 is a strong-lined quasar at $z=1.665$ with
$M_{\rm B}=-24.7$. The lines shown in Figure~5 have FWHM of
$\approx 5000$~km~s$^{-1}$, and the optical continuum slope
of $\alpha=-1.2\pm 0.4$ is consistent with that of `normal' 
blue quasars. 

Using our HET spectrum, we estimate $(V-R)\approx +0.3$  
for source~I3. For the other sources we estimate 
$(V-R)\approx +0.5$.


\subsection{The Blank-Field Sources}

We have compared the properties of the blank-field sources to those of
the other sources to gain clues to their nature.
Examination of Figure~6 shows that the blank-field sources
are not the faintest 2--8~keV sources in our sample; we have 
obtained successful HET identifications for sources with comparable 
or smaller 2--8~keV fluxes. This is comforting in that it
suggests that our blank-field sources are indeed reliable X-ray
detections. 
Figures~6a and 6b also show that the blank-field sources
have larger X-ray to $V$-band flux ratios than the other sources, as
expected. However, these X-ray to $V$-band flux ratios are still
consistent with those expected for AGN (compare with 
Figure~1 of Maccacaro et~al. 1988). Figure~6c suggests 
that the blank-field sources may be somewhat harder than 
the other sources, but we do not consider this result to be 
statistically significant at present. 

If the blank-field X-ray sources are in normal $L^\star$ galaxies
(Kirshner et~al. 1983; Efstathiou, Ellis \& Peterson 1988),
they must be at moderately high redshifts to explain their 
nondetections in the INT image (corresponding to $R>21.7$). 
To avoid detection in the INT image, an $L^\star$ Scd galaxy 
must have $z\simgt 0.75$, and an $L^\star$ elliptical, 
$z\simgt 0.55$. Typical $L^\star$ galaxies 
would thus need to be at higher redshifts than that of the 
cluster \crsscluster. 
The moderately high redshifts required for host galaxies also serve to 
rule out single extragalactic X-ray binaries and other low-luminosity 
X-ray sources associated with galaxies from creating the observed X-ray 
emission (unless the host galaxies have extremely low optical luminosities;
we would have detected a host galaxy that is sub-$L^\star$ by two
magnitudes to $z=0.2$). 
Large hard X-ray luminosities of 
$L_{2-8}\simgt 5\times 10^{41}$~erg~s$^{-1}$ are 
required for $z\simgt 0.2$. Such hard 
X-ray luminosities are commonly seen among local AGN. 
They might also be generated by the most extreme `pure' starburst 
galaxies, although even the most X-ray luminous starbursts known at
present have hard X-ray luminosities 
$\simlt 3\times 10^{41}$~erg~s$^{-1}$
(e.g., Moran, Lehnert \& Helfand 2000). 
In addition, the X-ray spectra of the blank-field sources are
significantly harder than those seen for `pure' starbursts
at low redshift. 

If the blank-field X-ray sources are `bona-fide' quasars, with 
$M_{\rm B}<-22.3$ for our adopted cosmology, they would need to 
have $z>1.75$ to avoid detection on the INT image.
Quasars with $z<3$ and $M_{\rm B}<-23.2$ would have been seen in the 
INT image, and a quasar as luminous as 3C273 would have been 
detected at redshifts greater than 4. 


\section{Discussion and Conclusions}

Our results, obtained from a small but well-defined sample of 
2--8~keV sources, extend previous X-ray 
background studies in several ways. First, 
we have detected and securely identified hard X-ray sources about 
an order of magnitude fainter than has previously been possible. 
At our 2--8~keV flux limit of 
$\approx$~(4--8)$\times 10^{-15}$~erg~cm$^{-2}$~s$^{-1}$
(spatially dependent; see \S2.2), we find 
ten sources (including AG1) in our 63.5~arcmin$^2$ search 
area, corresponding to a 2--8~keV source density of $570\pm 180$~deg$^{-2}$.
Even allowing for the possibility of one spurious source detection, 
this source density is still $\approx 10$ times larger 
than previous number counts in this energy band (e.g., Ogasaka et~al. 1998;
Giommi et~al. 1998), and down to $\approx 2\times 10^{-14}$~erg~cm$^{-2}$~s$^{-1}$ 
our number counts appear consistent with the \asca\ fluctuation analyses
of Gendreau et~al. (1998). However, our source counts
may be somewhat enhanced due to the presence of the cluster
\crsscluster\ (see below). The fact that we detect sources down to 
$\approx 4\times 10^{-15}$~erg~cm$^{-2}$~s$^{-1}$ suggests that the 
number counts versus flux relation  
departs from the Euclidean form (the X-ray background
would be resolved at $\approx 10^{-14}$~erg~cm$^{-2}$~s$^{-1}$ without
a break in the $\log N$--$\log S$ slope), although further data are 
clearly needed to quantify the break parameters. 

Four of our five S3 sources with optical spectroscopy have
$z<0.3$, which clearly exclude them from being associated with the cluster 
\crsscluster\ at $z=0.516$. The fifth, 
the QSO \crssqso, differs in redshift from the 
cluster by $\Delta z=0.023$ corresponding to a line-of-sight separation 
of $\approx 100$~Mpc. While this QSO is certainly not a bound member of the 
cluster, it could be associated with the large-scale cosmic structure producing 
the cluster. As discussed in \S3.3, if any of our blank-field sources 
(or AG1) lie in the cluster, they must be sub-$L^\star$ galaxies 
producing large hard X-ray luminosities 
of $L_{2-8}\simgt 4\times 10^{42}$~erg~s$^{-1}$.
Even in the most conservative (and unlikely) case, where we allow 
all objects with unknown $z$ to lie in the cluster, our
cluster-corrected source density is still a factor of $\approx 4.5$ 
times higher than previously attained by \asca\ and \sax. The same
statement obtains for possible gravitational lensing effects by the
cluster. 

We detect nine of our ten 2--8~keV sources in the 0.2--2~keV band. 
While our statistics are admittedly limited, this result is consistent 
with the finding by Giommi, Fiore \& Perri (1998) that most hard X-ray 
sources have soft X-ray counterparts, and it extends this result downward 
in flux by about an order of magnitude (see \S1 for discussion). Down to 
our flux limit, we can show with $>90$\% confidence that hard-band only 
sources comprise $<40$\% of the total hard-band source population. 
Deeper \chandra\ observations are needed to determine if a large 
population of hard-band only sources emerges at still fainter flux levels. 


\acknowledgments

We thank 
C.S.~Crawford and A.C.~Fabian for providing the archival INT image, 
G.M.~Hill and M.~Shetrone for help with the HET data acquisition, 
P.S.~Broos, A.C.~Fabian, E.D.~Feigelson and G.~Hasinger for helpful discussions, and
D.H.~Saxe for valuable computer support.  
We thank all the members of the \chandra\ team for their enormous efforts. 
We gratefully acknowledge the financial support of 
NASA grant NAS~8-38252 (GPG, PI), 
NASA LTSA grant NAG5-8107 (WNB),
NASA GSRP grant NGT5-50247 (AEH), and  
NSF grant AST99-00703~(DPS). 
The HET is a joint project of 
the University of Texas at Austin,
the Pennsylvania State University,  
Stanford University,
Ludwig-Maximillians-Universit\"at M\"unchen, and 
Georg-August-Universit\"at G\"ottingen.  
The HET is named in honor of its principal benefactors,
William P. Hobby and Robert E. Eberly.  
The Marcario LRS is a joint project of 
the University of Texas at Austin, 
the Instituto de Astronomia de la Universidad Nacional Autonoma de Mexico, 
Ludwig-Maximillians-Universit\"at M\"unchen, 
Georg-August-Universit\"at G\"ottingen,
Stanford University, and 
the Pennsylvania State University. 
This research is partially based upon data from the Isaac Newton Group 
archive.

\clearpage
 
 
\begin{deluxetable}{lcccccccc}
\tablecaption{Basic X-ray Properties of the 2--8~keV Sources. \label{tbl-1}}
\tablewidth{0pt}
\scriptsize
\tablehead{
\colhead{Source} & 
\colhead{X-ray} & 
\colhead{X-ray} &
\colhead{Aim point} &
\colhead{0.2--2~keV} & 
\colhead{2--8~keV} &
\colhead{Hard/Soft} &
\colhead{0.2--2~keV} & 
\colhead{2--8~keV} \\
\colhead{Name} & 
\colhead{$\alpha_{2000}$} & 
\colhead{$\delta_{2000}$} &
\colhead{distance ($^\prime$)} & 
\colhead{counts} & 
\colhead{counts} &
\colhead{ratio} &
\colhead{significance} &
\colhead{significance} 
}
\startdata
1   & 00$^{\rm h}$ 30$^{\rm m}$ 26.0$^{\rm s}$ & +26$^{\circ}$ 16$^{\prime}$ 48.9$^{\prime\prime}$ & 2.0 & $128.4\pm 11.5$  & $9.4\pm 3.2$    & $0.073\pm 0.026$ & 41.8     & 4.4        \nl
2   & 00$^{\rm h}$ 30$^{\rm m}$ 27.8$^{\rm s}$ & +26$^{\circ}$ 15$^{\prime}$ 14.9$^{\prime\prime}$ & 3.1 & $36.4\pm 6.6$    & $9.2\pm 3.6$    & $0.25\pm 0.11$   & 6.8      & 3.0        \nl
3   & 00$^{\rm h}$ 30$^{\rm m}$ 30.8$^{\rm s}$ & +26$^{\circ}$ 16$^{\prime}$ 00.0$^{\prime\prime}$ & 2.2 & $30.8\pm 5.7$    & $12.0\pm$ 3.6   & $0.39\pm 0.14$   & 11.3     & 5.2        \nl
4   & 00$^{\rm h}$ 30$^{\rm m}$ 33.3$^{\rm s}$ & +26$^{\circ}$ 14$^{\prime}$ 50.9$^{\prime\prime}$ & 3.3 & $34.9\pm 6.1$    & $6.2\pm 2.6$    & $0.18\pm 0.08$   & 13.1     & 2.8        \nl
5   & 00$^{\rm h}$ 30$^{\rm m}$ 34.7$^{\rm s}$ & +26$^{\circ}$ 16$^{\prime}$ 29.8$^{\prime\prime}$ & 1.8 & $<7.5$           & $8.1\pm 3.0$    & $>0.68$          & $\cdots$ & 3.6        \nl
6   & 00$^{\rm h}$ 30$^{\rm m}$ 39.5$^{\rm s}$ & +26$^{\circ}$ 20$^{\prime}$ 54.9$^{\prime\prime}$ & 3.1 & $1830.5\pm 43.0$ & $413.1\pm 20.4$ & $0.23\pm 0.01$   & 355.3    & 122.4      \nl
7   & 00$^{\rm h}$ 30$^{\rm m}$ 42.6$^{\rm s}$ & +26$^{\circ}$ 17$^{\prime}$ 46.1$^{\prime\prime}$ & 2.3 & $21.9\pm 4.9$    & $11.9\pm 3.6$   & $0.54\pm 0.23$   & 8.1      & 5.1        \nl
8   & 00$^{\rm h}$ 30$^{\rm m}$ 47.8$^{\rm s}$ & +26$^{\circ}$ 16$^{\prime}$ 47.0$^{\prime\prime}$ & 3.7 & $141.9\pm 12.3$  & $27.8\pm 5.7$   & $0.20\pm 0.04$   & 34.3     & 8.0        \nl
9   & 00$^{\rm h}$ 30$^{\rm m}$ 51.3$^{\rm s}$ & +26$^{\circ}$ 17$^{\prime}$ 11.9$^{\prime\prime}$ & 4.3 & $99.4\pm 10.4$   & $42.1\pm$ 6.9   & $0.42\pm 0.09$   & 23.5     & 11.8       \nl
\tablevspace{0.2cm}
AG1 & 00$^{\rm h}$ 30$^{\rm m}$ 41.6$^{\rm s}$ & +26$^{\circ}$ 17$^{\prime}$ 40.8$^{\prime\prime}$ & 2.1 & $65.9\pm 8.3$    & $13.1\pm 4.0$   & $0.20\pm 0.07$   & 22.4     & 4.5        \nl
\tablevspace{0.2cm}
I3  & 00$^{\rm h}$ 31$^{\rm m}$ 08.4$^{\rm s}$ & +26$^{\circ}$ 20$^{\prime}$ 56.3$^{\prime\prime}$ & 8.5 & $136.4\pm 12.4$  & $36.5\pm 6.9$   & $0.27\pm 0.06$   & 26.4     & 8.4        \nl
\enddata
\tablenotetext{}{Sources~1--9 are located on the S3 CCD and source~I3 is located on the I3 CCD; properties
for these sources have been found using filtering on ACIS grades 0, 2, 8, 16 and 64. 
Source~AG1 is only found when \asca\ grade filtering is used (see \S2.4); properties for this
source have been found using filtering on \asca\ grades 0, 2, 3, 4 and 6. 
%
%
Note that we have only listed the sources that are detected in the 2--8~keV band. 
}
\end{deluxetable}




\begin{deluxetable}{lcccccl}
\tablecaption{X-ray Fluxes/Luminosities and Optical Properties of the 2--8~keV Sources. \label{tbl-2}}
\tablewidth{0pt}
\scriptsize
\tablehead{
\colhead{Source} & 
\colhead{0.2--2~keV flux} & 
\colhead{2--8~keV flux} &
\colhead{2--8~keV $L_{\rm X}$} &
\colhead{} &
\colhead{} &
\colhead{} \\
\colhead{Name} & 
\colhead{(erg cm$^{-2}$ s$^{-1}$)} & 
\colhead{(erg cm$^{-2}$ s$^{-1}$)} &
\colhead{(erg s$^{-1}$)} &
\colhead{$R$} &
\colhead{$z$} &
\colhead{Notes$^{\rm a}$}
}
\startdata
1         & $1.0\times 10^{-14}$  & $6.0\times 10^{-15}$ & $1.1\times 10^{42}$           & 18.4     & 0.269            & HET                  \nl
2         & $2.9\times 10^{-15}$  & $5.9\times 10^{-15}$ & $\cdots$                      & $>21.7$  & $\cdots$         & BF                   \nl
3         & $2.5\times 10^{-15}$  & $7.7\times 10^{-15}$ & $\cdots$                      & $>21.7$  & $\cdots$         & BF                   \nl
4         & $2.9\times 10^{-15}$  & $4.1\times 10^{-15}$ & $6.1\times 10^{41}$           & 18.7     & 0.247            & HET                  \nl
5         & $<5.9\times 10^{-16}$ & $5.2\times 10^{-15}$ & $\cdots$                      & $>21.7$  & $\cdots$         & BF                   \nl
6         & $1.6\times 10^{-13}$  & $3.3\times 10^{-13}$ & $2.1\times 10^{44}$           & 16.9     & 0.493            & \crssqso\ (QSO)      \nl
7         & $1.7\times 10^{-15}$  & $7.7\times 10^{-15}$ & $\cdots$                      & $>21.7$  & $\cdots$         & BF                   \nl
8         & $1.1\times 10^{-14}$  & $1.8\times 10^{-14}$ & $2.7\times 10^{42}$           & 18.5     & 0.247            & \crsssy2 (Seyfert~2) \nl
9         & $8.0\times 10^{-15}$  & $2.9\times 10^{-14}$ & $1.1\times 10^{42}$$^{\rm b}$ & 19.1     & 0.129$^{\rm b}$  & HET                  \nl
\tablevspace{0.2cm}
AG1       & $4.2\times 10^{-15}$  & $5.5\times 10^{-15}$ & $\cdots$                      & 21.5     & $\cdots$         & \asca\ grade source  \nl
\tablevspace{0.2cm}
I3        & $1.9\times 10^{-14}$  & $2.0\times 10^{-14}$ & $1.9\times 10^{44}$           & 19.1     & 1.665            & HET \nl
\enddata
\tablenotetext{a}{Sources noted as `HET' are those for which we present HET 
spectra, and sources noted as `BF' (for `Blank Field') are those without optical 
counterparts.}
\tablenotetext{b}{We consider this redshift and luminosity to be only 
tentative (see \S3.2 for details).}
\end{deluxetable}

\clearpage


%
%


%
%


\clearpage


\plotfiddle{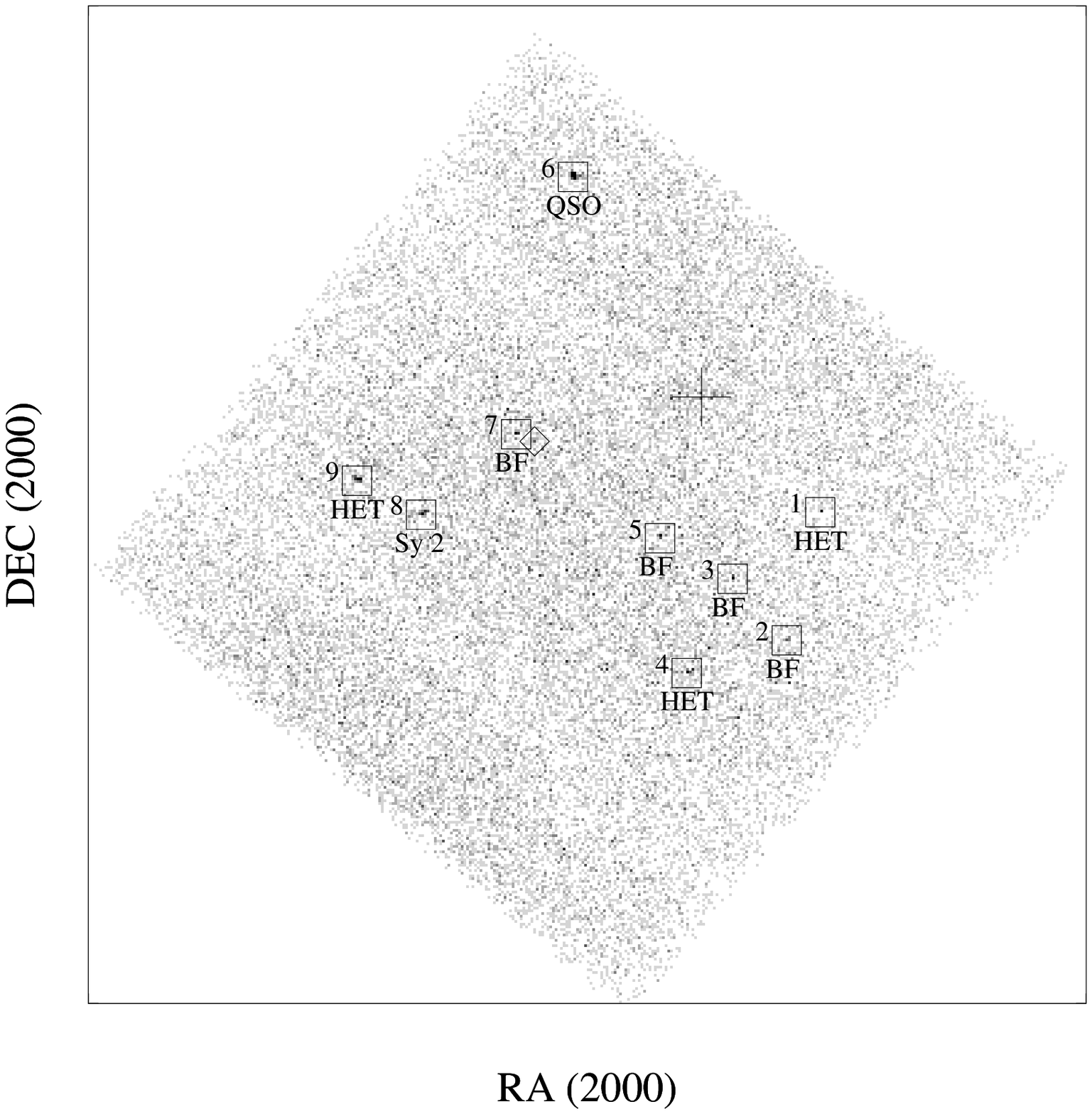}{6.8in}{0.0}{80.0}{80.0}{-280.0}{-20.0}
\vspace{-4cm}
\figcaption[]
{Image of the ACIS S3 CCD from 2--8~keV with the {\sc wavdetect} sources marked 
and numbered. The box sizes are arbitrary ($\approx 20^{\prime\prime}$ on a 
side); the positional accuracy is much better than the box size. 
The cross shows the position of the aim point and the 
cluster \crsscluster\ ($z=0.516$), and the diamond 
near source~7 shows the position of source~AG1 (see \S2.4). 
The QSO \crssqso\ ($z=0.493$) and the Seyfert~2 \crsssy2\ ($z=0.247$) 
are labeled as `QSO' and `Sy~2', respectively. 
Sources for which we present HET spectra are labeled `HET' and sources
without optical counterparts are labeled `BF' (for `Blank Field'). 
North is at the top, and East is to the left. 
For scale, the CCD is $8.3^{\prime}$ on a side. 
\label{fig1}}

\clearpage


\plotfiddle{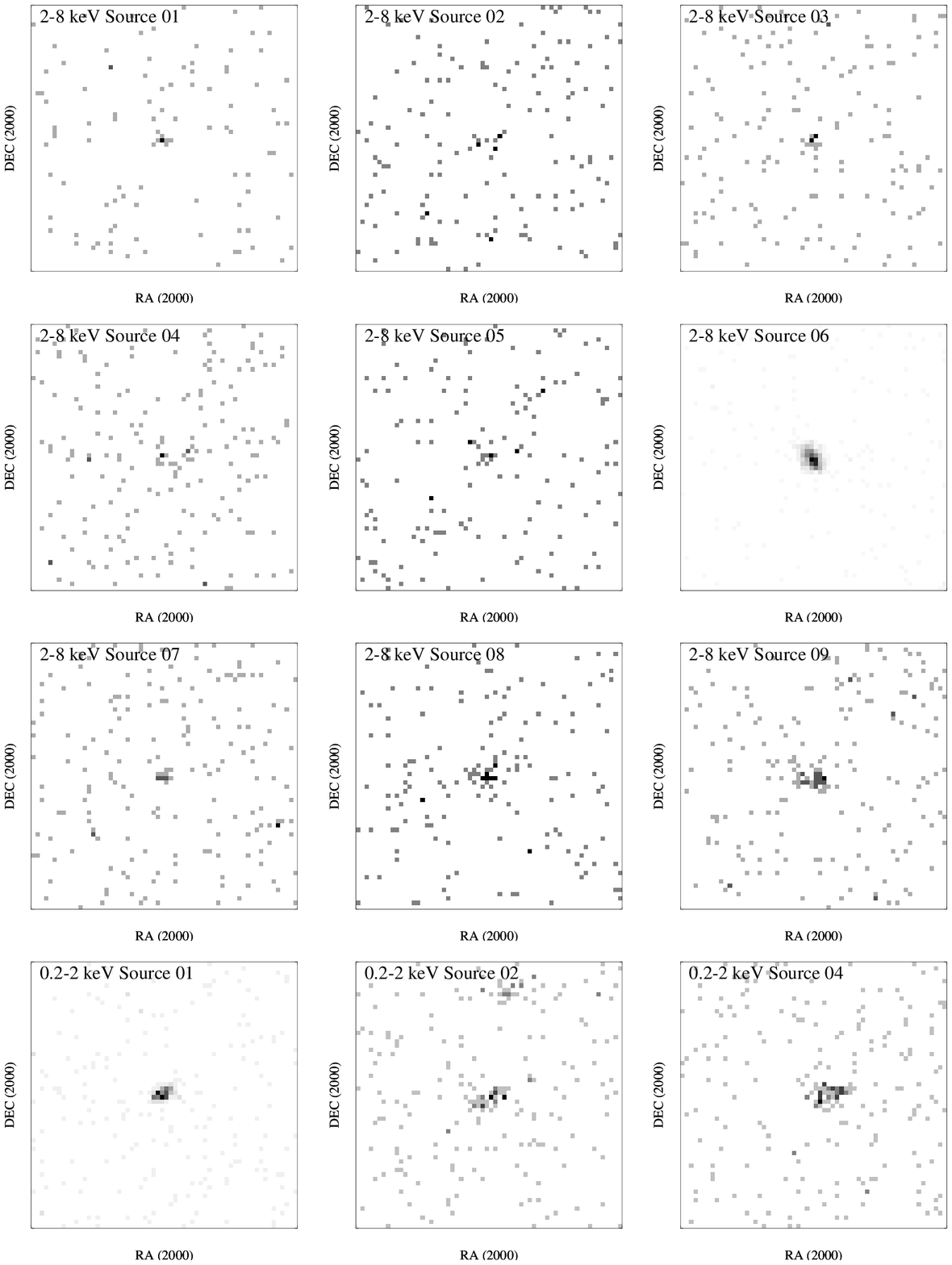}{6.8in}{0.0}{80.0}{80.0}{-260.0}{-60.0}
\figcaption[]
{ACIS images of the 2--8~keV sources from the S3 CCD. Each
image is centered on the corresponding position from Table~1, 
and each is $30^{\prime\prime}$ on a side. Each pixel is
$0.5^{\prime\prime}$ on a side. To further demonstrate the 
reality of sources 1, 2 and 4, we also show their independent 
0.2--2~keV images. The grayscale levels are linear and 
vary from image to image, but white corresponds to zero in 
all images. Most of the non-white pixels away from the sources 
themselves have one count. 
\label{fig2}}

\clearpage


\plotfiddle{brandt.fig3.ps}{6.8in}{-90}{80.0}{80.0}{-260.0}{+550.0}
\vspace{-3 cm}
\figcaption[]
{Plot of the hard-band to soft-band ratio versus column density at 
$z=0$ for power-law models with photon indices of
$\Gamma=1.7$ (dashed curve),
$\Gamma=1.9$ (solid curve) and
$\Gamma=2.1$ (dot-dashed curve).
The data points show the band 
ratios for sources 3, 5, 7 and 9 (the hardest sources we find). 
These sources have been arbitrarily placed on the $\Gamma=1.9$
curve; better X-ray spectra would be needed to determine their
underlying photon indices. Provided these 
sources have intrinsic X-ray continua that are similar to those
of Seyferts and QSOs, they appear to have column densities larger than 
a few times $10^{21}$~cm$^{-2}$. Source~5 is likely to have a column 
density $\simgt 10^{22}$~cm$^{-2}$. Note that the probable absorption for 
these sources is presumably intrinsic, and thus correcting for the 
effects of redshift will only increase the inferred column densities. 
The curves have been calculated assuming only ACIS grade
0, 2, 8, 16 and 64 events are used. 
\label{fig3}}

\clearpage


\plotfiddle{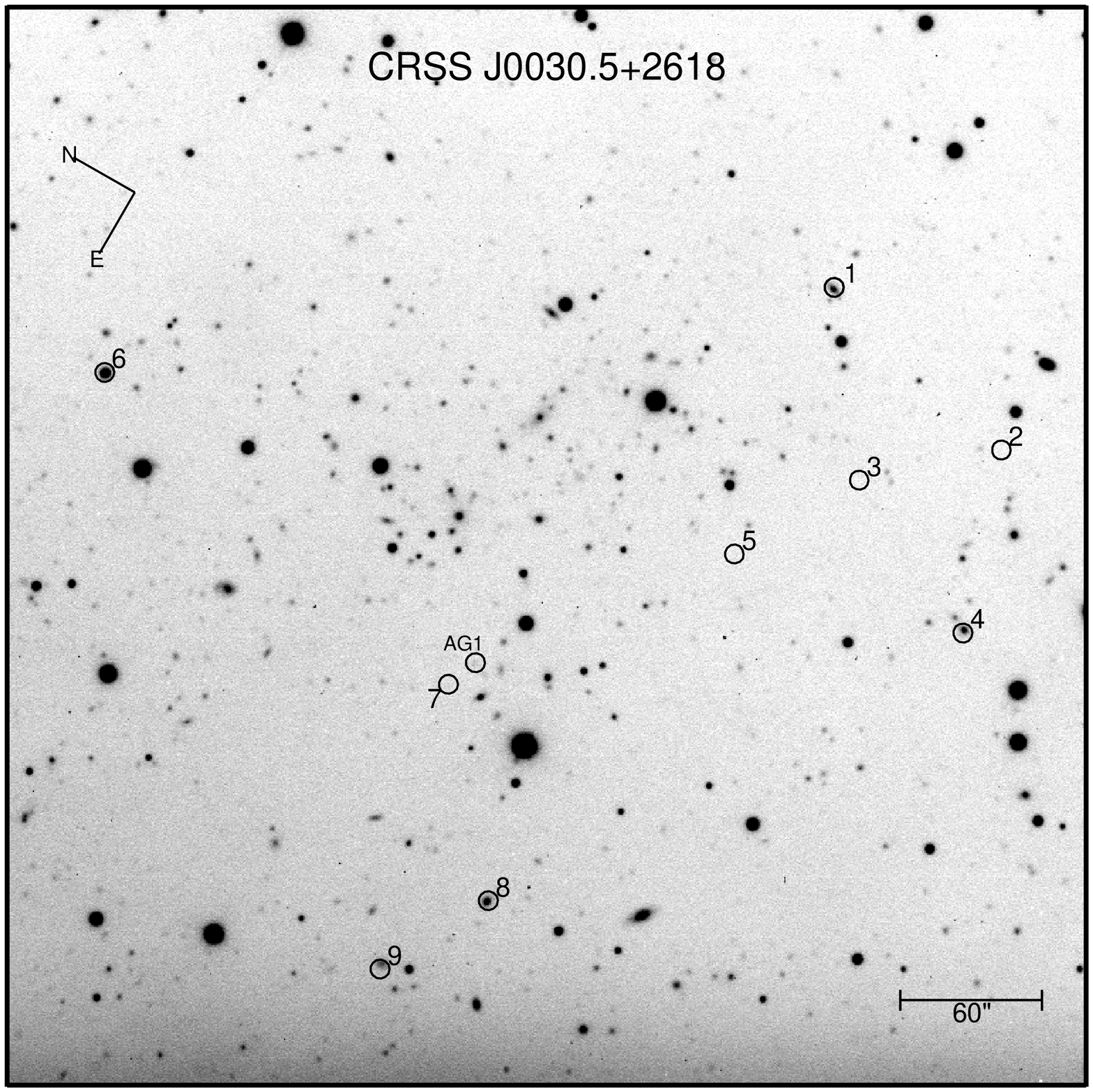}{6.8in}{0.0}{70.0}{70.0}{-220.0}{40.0}
\vspace{-1 cm}
\figcaption[]
{INT image of the field of the cluster \crsscluster. The field
is $7.5^{\prime}$ on a side. We have marked our \chandra\ sources
with circles of $4^{\prime\prime}$ radius (note that this is somewhat
larger than our positional uncertainty). The cluster is located
about $1^\prime$ North of source~5. 
\label{fig4}}

\clearpage


\plotfiddle{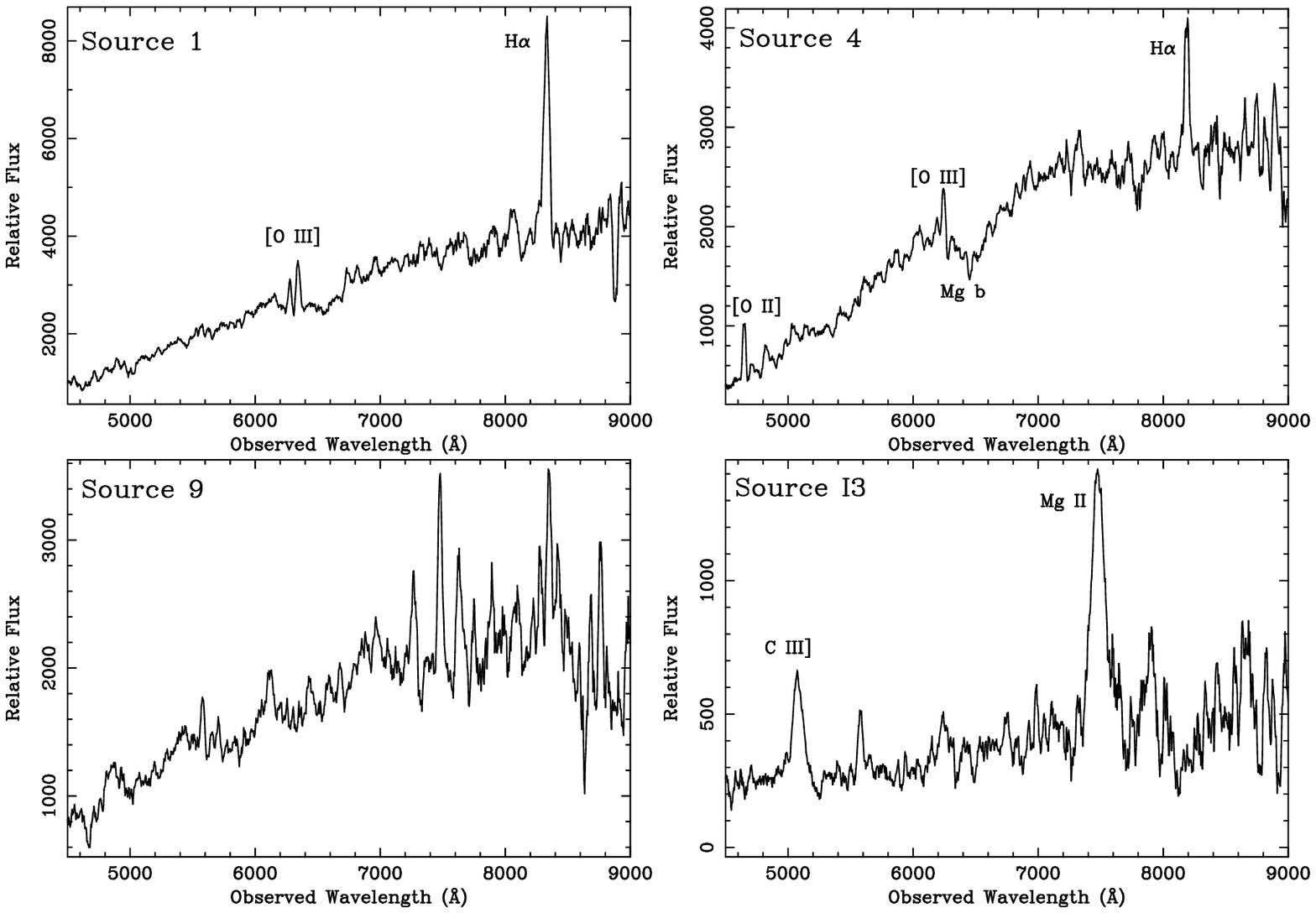}{6.8in}{0.0}{90.0}{90.0}{-280.0}{-40.0}
\vspace{-4 cm}
\figcaption[]
{HET LRS spectra for the \chandra\ sources 1, 4, 9 and I3. The spectral
resolution is 24~\AA.
\label{fig5}}

\clearpage


\plotfiddle{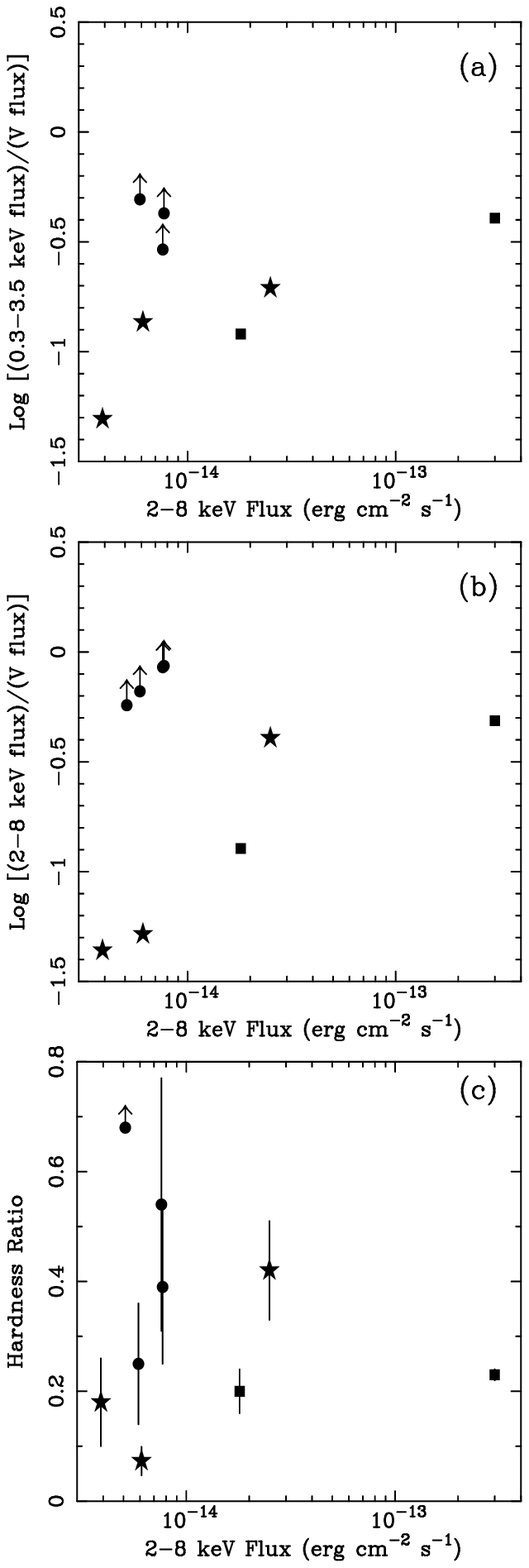}{6.8in}{0.0}{80.0}{80.0}{-240.0}{-40.0}
\vspace{-2 cm}
\figcaption[]
{X-ray to $V$-band flux ratios and X-ray band ratios for our sources. 
Stars indicate sources with HET spectra, squares indicate CRSS
sources, and solid dots indicate blank-field sources. 
Source~5 is not plotted in panel (a) because it does not have a 
soft-band detection, and two of the solid dots overlap in panel (b). 
The fluxes used in panels (a) and (b) are corrected for absorption
by the Galaxy. Panel (a) is plotted for 0.3--3.5~keV for comparison
with the X-ray to $V$-band flux ratios given by Stocke et~al. (1991) 
and Schmidt et~al. (1998). Our X-ray to $V$-band flux ratios are
on average somewhat smaller than those found by these authors, probably a
selection effect due to our sensitive X-ray data and our relatively
shallow optical data. 
\label{fig6}}

\clearpage


\end{document}